\documentclass[sigconf,screen]{acmart}
\usepackage{tabularx}
\usepackage{enumitem}
\usepackage{listings}
\usepackage{url}
\usepackage{float}
\usepackage{subcaption}
\usepackage{enumerate}
\usepackage{threeparttable}
\usepackage{multirow,multicol}

\setcopyright{acmlicensed}
\acmDOI{10.1145/3663529.3663801}
\acmYear{2024}
\copyrightyear{2024}
\acmSubmissionID{fsecomp24demo-p31-p}
\acmISBN{979-8-4007-0658-5/24/07}
\acmConference[FSE Companion '24]{Companion Proceedings of the 32nd ACM International Conference on the Foundations of Software Engineering}{July 15--19, 2024}{Porto de Galinhas, Brazil}
\acmBooktitle{Companion Proceedings of the 32nd ACM International Conference on the Foundations of Software Engineering (FSE Companion '24), July 15--19, 2024, Porto de Galinhas, Brazil}
\received{2024-01-29}
\received[accepted]{2024-04-15}

\AtBeginDocument{%
  }

\begin{document}

\title{ChatUniTest: A Framework for LLM-Based Test Generation}

\author{Yinghao Chen}
\affiliation{%
  \institution{Zhejiang University}
  \country{China}}
\email{yh_ch@zju.edu.cn}

\author{Zehao Hu}
\affiliation{%
  \institution{Zhejiang University}
  \country{China}}
\email{Huzehao@zju.edu.cn}

\author{Chen Zhi}
\authornote{Chen Zhi is the corresponding author.}
\affiliation{%
  \institution{Zhejiang University}
  \country{China}}
\email{zjuzhichen@zju.edu.cn}

\author{Junxiao Han}
\affiliation{%
  \institution{Hangzhou City University}
  \country{China}}
\email{hanjx@hzcu.edu.cn}

\author{Shuiguang Deng}
\affiliation{%
  \institution{Zhejiang University}
  \country{China}}
\email{dengsg@zju.edu.cn}

\author{Jianwei Yin}
\affiliation{%
  \institution{Zhejiang University}
  \country{China}}
\email{zjuyjw@cs.zju.edu.cn}

\renewcommand{\shortauthors}{Chen et al.}

\begin{abstract}

Unit testing is an essential yet frequently arduous task. Various automated unit test generation tools have been introduced to mitigate this challenge. Notably, methods based on large language models (LLMs) have garnered considerable attention and exhibited promising results in recent years. Nevertheless, LLM-based tools encounter limitations in generating accurate unit tests. This paper presents ChatUniTest, an LLM-based automated unit test generation framework. ChatUniTest incorporates an adaptive focal context mechanism to encompass valuable context in prompts and adheres to a generation-validation-repair mechanism to rectify errors in generated unit tests.
Subsequently, we have developed ChatUniTest Core, a common library that implements core workflow, complemented by the ChatUniTest Toolchain, a suite of seamlessly integrated tools enhancing the capabilities of ChatUniTest. Our effectiveness evaluation reveals that ChatUniTest outperforms TestSpark and EvoSuite in half of the evaluated projects, achieving the highest overall line coverage.
Furthermore, insights from our user study affirm that ChatUniTest delivers substantial value to various stakeholders in the software testing domain. 
ChatUniTest is available at \url{https://github.com/ZJU-ACES-ISE/ChatUniTest}, and the demo video is available at \url{https://www.youtube.com/watch?v=GmfxQUqm2ZQ}.

\end{abstract}

%




\maketitle

\section{introduction}

Unit testing is a critical practice in software development, ensuring the quality of software applications. However, manually writing and maintaining high-quality unit tests can be a daunting task. 
Automated unit test generation tools have been introduced to alleviate the burden on developers of writing essential unit tests. Existing tools~\cite{evosuite,randoop,Kex,UTBot,TestMe} primarily rely on program analysis techniques for the generation of unit tests. 
For instance, EvoSuite~\cite{evosuite}, a notable search-based tool, demonstrates effectiveness in achieving reasonable coverage. However, a common drawback is that tests generated by these tools often lack explainability and readability. 

Recently, the exploration of deep learning techniques, particularly leveraging large language models (LLMs), for the generation of unit tests has shown significant promise. TestPilot~\cite{TestPilot}, a leading LLM-based tool developed by GitHub, introduces an adaptive test generation mechanism based on Codex~\cite{codex}. TestSpark~\cite{TestSpark}, a JetBrains IDEA plugin, provides dual modes for test generation: one leverages OpenAI and JetBrains' AI for LLM-based generation, and the other uses EvoSuite for search-based generation.

However, we have identified two primary limitations associated with llm-based tools. Firstly, the constraint on context length imposes a limitation on the capacity of LLMs to process all relevant information and generate a comprehensive unit test. Although recent LLMs exhibit the ability to handle longer contexts as input, it remains imperative to provide concise and precise context to LLMs for reasons of economic cost and the lost-in-the-middle effect~\cite{lostinTheMiddle}. Secondly, due to insufficient validation mechanisms, LLMs often produce incorrect tests with various errors, such as \textit{"cannot find symbol"} during compilation and \textit{"AssertionFailedError"} during runtime, which require considerable effort for manual repair. 


In this paper, we introduce ChatUniTest, a framework to generate unit tests automatically based on LLMs. ChatUniTest employs an adaptive focal context generation mechanism to overcome the first limitation and implements a generation-validation-repair mechanism to tackle the second. These enhancements significantly improve LLM’s effectiveness in generating unit tests.
The main contributions of our work are as follows:
\begin{itemize}
    \item We present ChatUniTest, a unit test generation framework based on LLM. By utilizing innovative mechanisms such as adaptive focal context and generation-validation-repair mechanisms, we have developed the \textbf{ChatUniTest Core}~\cite{chatunitest-core} to assist researchers and tool builders. 
    \item We have implemented the \textbf{ChatUniTest Toolchain}, which includes the Maven plugin~\cite{chatunitest-maven-plugin} and the IntelliJ IDEA plugin~\cite{chatunitest-idea-plugin}. It offers convenience to users and caters to the requirements of various scenarios. 
\end{itemize}

\section{Related Work}


The research history of automated unit test generation spans several decades. However, traditional methods~\cite{evosuite,randoop} for generating unit tests exhibit significant deficiencies in terms of explainability and readability. Recently, LLMs have demonstrated remarkable performance across various programming tasks. In response, researchers have proposed numerous test generation approaches\cite{athenatest,a3test,TestPilot,catlm,ChatTester,SysPrompt,TestSpark,COVERUP,TELPA,wang2024software} based on these models, which are capable of generating well-explained and readable unit tests. Notably, TestPilot\cite{TestPilot} and TestSpark\cite{TestSpark} are typical test generation solutions based on LLMs, and they provide user-friendly services or tools for this purpose.
TestPilot\cite{TestPilot} automates unit test generation for JavaScript programs. It utilizes function signatures, implementation, and usage examples extracted from documentation. However, its reliance on documentation limits its applicability to undocumented or sparsely documented programs.
TestSpark\cite{TestSpark} introduces a novel approach to unit test generation within the IntelliJ IDEA. Leveraging both search-based test generation tools and large language models (LLMs), TestSpark employs a feedback cycle between the IDE and LLMs, enhancing test generation effectiveness. 

\section{Approach}

\begin{figure}[t]
  \includegraphics[width=\linewidth]{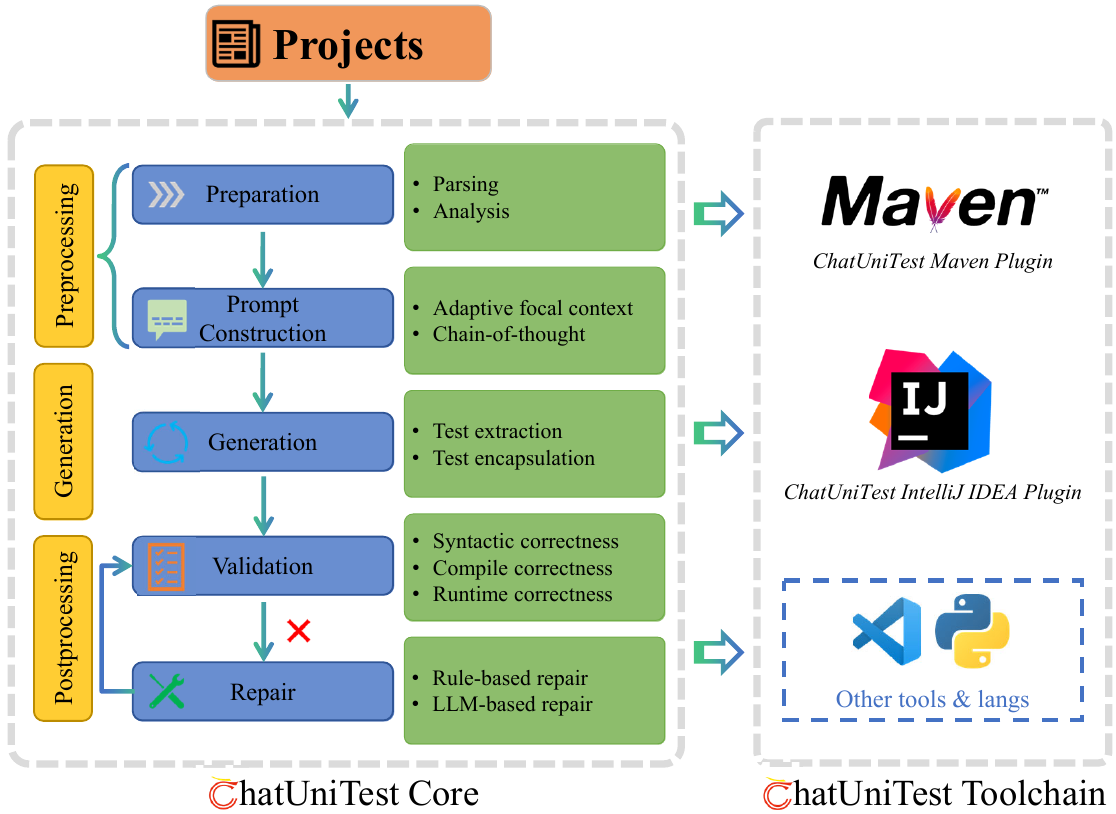}
  \setlength{\abovecaptionskip}{-2mm}
  \setlength{\belowcaptionskip}{-2mm}
  \caption{overview of chatunitest-core and tools derived}
  \Description{overview of chatunitest-core and tools derived}
  \label{fig:chatunitest}
  \vspace*{-3mm}
\end{figure}


As shown in Figure \ref{fig:chatunitest}, \textbf{ChatUniTest Core} establishes the overall architecture designed for both researchers and tool builders. This architecture encompasses three phases and five steps: preprocessing (including preparation and prompt construction), generation, and postprocessing (including validation and repair). The preprocessing phase mainly focuses on how to unlock the capabilities of LLM, such as designing efficient and effective prompts. The postprocessing phase primarily focuses on how to mitigate the limitations of LLM, such as by repairing generated test cases.

\subsection{Preparation}

ChatUniTest initially requires preparation to optimize and accelerate generation. This includes parsing the target projects, analyzing dependencies, and obfuscating the code for privacy.

\paragraph{Parsing}
The primary objective of the parsing step is to extract raw information from the code. In this stage, ChatUniTest scans the project folder to identify all Java files. Each Java file is then converted into an Abstract Syntax Tree (AST). As ChatUniTest explores the AST, it collects information at both the class and method levels. For classes, it notes the package, imports, extends, implements, fields, and method signatures. For methods, it records the method signature, body, field access, getter/setter invocation, dependent class names, and method invocations.

\paragraph{Analysis}
The analysis step is primarily focused on leveraging static analysis techniques to extract deeper insights to promote unit test generation. For example, we intend to conduct program slicing to extract API call sample code and perform dependency analysis to build Object Construction Graphs (OCGs~\cite{OCG}), which can provide guidance for LLM to generate driver code. 


\subsection{Prompt Construction}

ChatUniTest employs an adaptive focal context generation mechanism and a prompt template to construct the input prompt for LLM. The adaptive focal context generation mechanism aimed to overcome the primary limitation. This mechanism formulates a context abundant in essential details about the focal method while excluding extraneous information from the focal class. The prompt template provides an approach to designing specific prompts for various requirements, or one can simply utilize the default template.

\paragraph{Adaptive Focal Context Generation}
The adaptive focal context generation mechanism is designed to add as much valuable information to the prompt as possible while ensuring that each addition stays within the token limit. This approach optimizes the focal method's context, facilitating the integration of the class context and the contexts of associated methods. 

The mechanism dynamically generates context according to the dependency relationships of the focal methods. Initially, the context encompasses the information of the focal method and the focal class. Based on the result of dependency analysis, other relevant information (such as signatures of dependent methods and classes) might also be integrated into the context. 

In the future, we plan to enhance the adaptive focal context generation mechanism. Our ultimate goal is to produce the optimal context according to the properties of each focal method, ensuring that LLMs have a clearer understanding, which in turn will elevate the quality of the generated tests.


In accordance with adaptive focal context generation, ChatUniTest initially parses the template file and substitutes the labels and placeholders with actual data. If the prompt tokens exceed the token limit, ChatUniTest will automatically eliminate the least relevant data in the user prompt, typically the last label or placeholder. By doing so, the prompt tokens are within the limit.



\paragraph{Prompt Template}

ChatUniTest utilizes the FreeMarker Java Template Engine\cite{freemarker} to generate the prompt based on the template and changing context. This approach, combined with the adaptive focal context mechanism, provides users with the flexibility to design diverse prompt templates tailored to their requirements while still reserving ample tokens for the LLM to produce helpful responses.



\paragraph{Chain-of-thought}
Based on the dependency graph constructed during preprocessing, we are implementing to create chain-of-thought (COT) prompts. Prior research~\cite{gtip} has already demonstrated that the COT can effectively enhance the capability of large language models (LLMs) in addressing software engineering tasks. We believe COT could make LLMs achieve more comprehensive understanding of the code, subsequently improves the quality of the generated tests.

\vspace*{-2mm}

\subsection{Generation}

Utilizing the generated focal context and the user-provided prompt template (or the default template if none is provided), the generation component integrates essential information in the context into the prompt. To guarantee sufficient tokens for test case generation, a fixed threshold is set to control how many tokens the prompt can consume. The context information is then filled within this threshold. Subsequently, the LLM is invoked using the LLM API.

Upon receiving a response, ChatUniTest utilizes its CodeExtractor component to extract tests from the response. Our experiments have shown that the LLM may provide either a single test method or an entire test class. To address this, we have implemented the TestSkeleton, which can encapsulate the test method automatically to construct a complete test class.
Once the test class is generated, it marks the completion of the test generation process.

\sloppy
Inspired by the CAT-LM~\cite{catlm}, we also build \textbf{ChatUniTest Models}\cite{chatunitest-models},  which provides fine-tuned models for Java test generation tasks based on \textbf{Code Llama}\cite{codellama}.

\vspace*{-2mm}

\subsection{Syntactic, Compile and Runtime Validation}

The extracted test will be forwarded to the validation component for further verification of its correctness. The validation process consists of three sub-steps: syntactic validation, compile validation, and runtime validation. A test is considered successfully generated only if it passes all the validation steps.

\paragraph{Syntactic Validation}
ChatUniTest first utilizes Java Parser to verify the syntax of generated tests. If a sytax error is detected at this stage, it can usually be fixed by the rule-based repair component. For example, an error \textit{"Parse error. Found <EOF>"} in JavaParser suggests that there might be a missing \textit{";"}, \textit{"\}"}, or another closing symbol in the code.

\paragraph{Compile Validation}
Following the syntactic validation, ChatUniTest proceeds to compile the test. This step not only checks for syntax errors but also examines certain semantic issues in the code. 
If the compile fails, the tool will then move on to the repair phase. For instance, the error \textit{"cannot find symbol"} suggests that there might be missing dependencies. It can usually be resolved with the appropriate import statement.

\paragraph{Runtime Validation}
After completing the compile validation, ChatUniTest begins the execution of the test. If any runtime errors are encountered, the tool will transition to the repair phase. As an example, the error \textit{"org.mockito.exceptions.misusing"} suggests that a mockito method is used incorrectly. Such as \textit{when()} method is used without providing a method call on a mock object as its argument. 

\subsection{Rule-based and LLM-based Repair}

If errors occur during the validation step, ChatUniTest initiates the repair process for rectification. There are two types of repair strategies: rule-based repair and LLM-based repair. The former are used to repair specific errors, and the latter are used as a fallback measurement to repair a broader range of unexpected errors.


\paragraph{Rule-based Repair}
ChatUniTest first employs its rule-based repair component to correct specific errors in the test. Conceptually, a repair rule consists of two parts: triggers and transformers. Triggers define when to apply the repair rule, and transformers formulate how to rewrite the code to resolve the error. Currently, ChatUniTest implements two simple repair rules.
\begin{itemize}[leftmargin=*]
    \item When the LLM's response exceeds the token limit, it can lead to truncated tests and syntactic errors. To address this, ChatUniTest first adds the necessary ending braces to validate the structure. If this fails, it searches for the \textit{"@Test"} annotation, truncates the code preceding it, and appends closing braces. If the repair fails or no tests remain post-removal, the process ends. 
    \item If a \textit{"cannot find symbol"} error arises during the compile validation process, ChatUniTest tries to address it by simply copying all import statements from the focal class into the generated tests. 
\end{itemize}

\paragraph{LLM-based Repair}
If errors persist in the generated test after applying all the rule-based repairs, it should be handed over to the LLM-based repair. Within this repair phase, ChatUniTest gathers details about the error type and message, integrates this with the incorrect test's context, and generates a prompt using a predefined template. If a test remains unresolved after a specified number of rounds, the entire process is terminated.

\section{Illustrative Example}
As shown in Figure \ref{fig:chatunitest}, we build the \textbf{ChatUniTest Toolchain} based on \textbf{ChatUniTest Core}. The toolchain provides plugins tailored for various scenarios and offers users a more streamlined experience. 
To demonstrate the capabilities of ChatUniTest, we examine a specific test case generated by the IntelliJ IDEA plugin.

\begin{figure}[H]
  \includegraphics[width=0.7\linewidth]{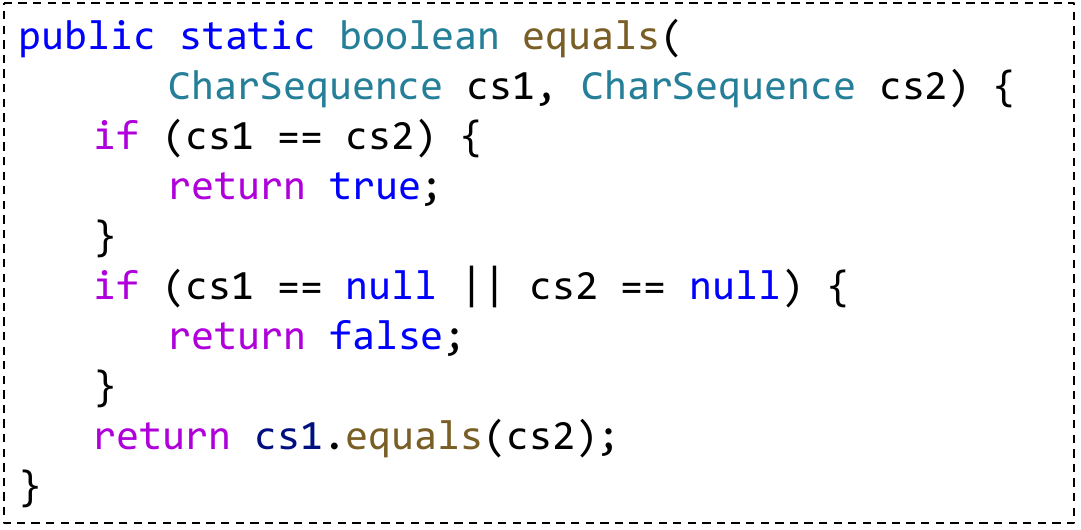}
  \caption{Buggy method \textit{equals} in Apache Commons.}
  \label{fig:example_b}
  \vspace*{-2mm}
\end{figure}

Consider the method \textit{equals} within the \textit{StringUtils} class, as depicted in Figure \ref{fig:example_b}. This method aims to determine if two given \textit{CharSequence} objects, \textit{cs1} and \textit{cs2}, are equal.

\begin{figure}[H]
  \includegraphics[width=\linewidth]{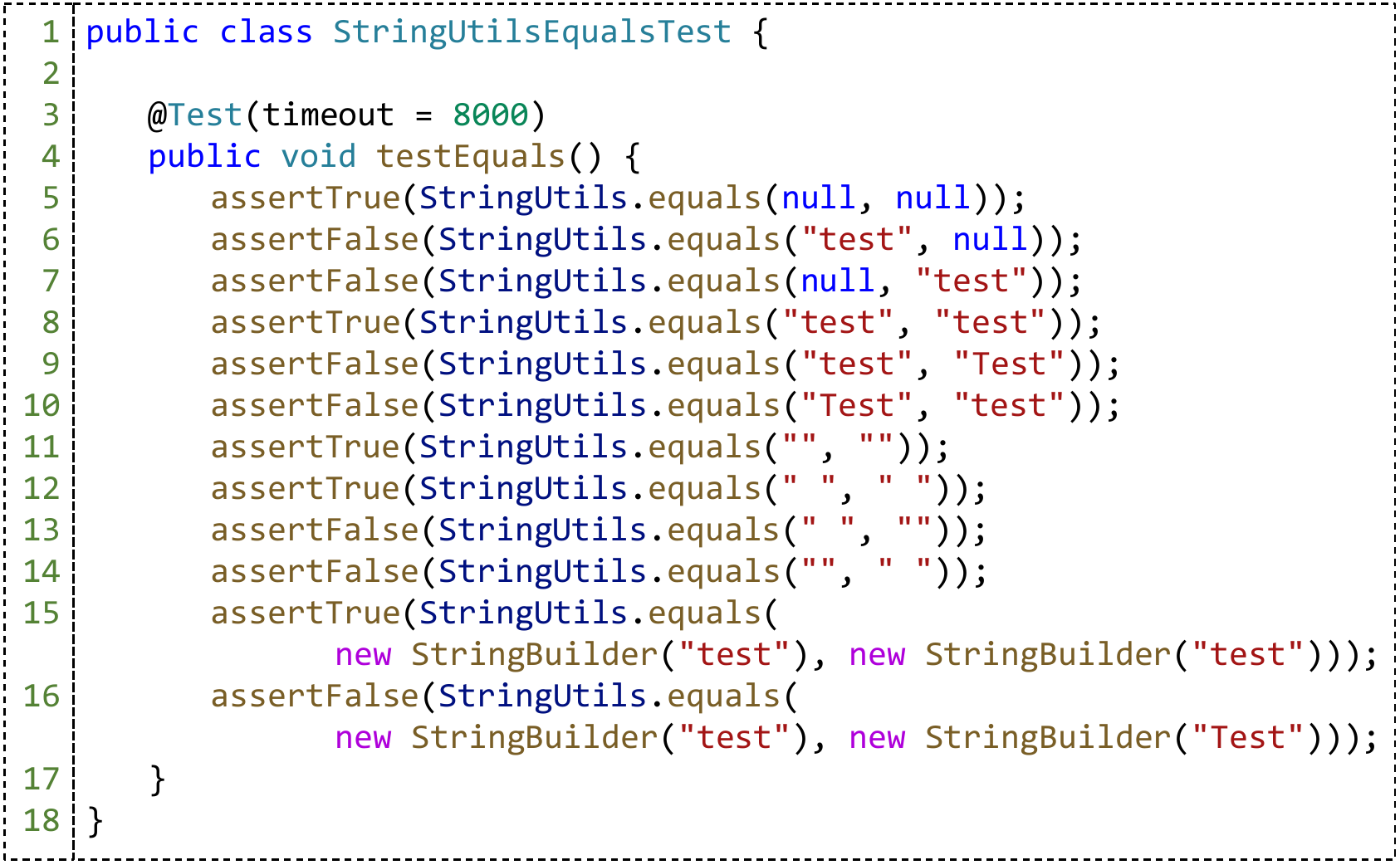}
  \setlength{\abovecaptionskip}{-2mm}
  \setlength{\belowcaptionskip}{-2mm}
  \caption{The Test Case generated by ChatUniTest.}
  \label{fig:result}
  \vspace*{-2mm}
\end{figure}

Developers can invoke ChatUniTest to generate unit tests for the method \textit{equals}. Figure \ref{fig:result} illustrates the generation results.
The generated test will throw \textit{"AssertionFailedError"} at line 15 because \textit{StringBuilder} implements \textit{CharSequence} but does not override the \textit{equals} method from the Object class. As a result, for \textit{StringBuilder}, the \textit{equals} method checks for reference equality rather than value equality. Therefore, the test detects a defect in the focal method.

\section{evaluation}

\paragraph{Effectiveness Evaluation}

Our initial evaluation aims to ascertain the effectiveness of ChatUniTest by assessing the quality of the tests it generates. To establish baselines for comparison, we selected two representative tools: EvoSuite (representing program-analysis-based tools) and TestSpark (representing LLM-based tools). Notably, TestPilot was excluded from this comparison due to its lack of support for Java.
In our experimental setup, we utilized the default configuration for EvoSuite and maintained consistent configurations between ChatUniTest and TestSpark. This included the use of gpt-3.5-turbo-0613 as the base model, with each round comprising one generation attempt followed by five repair attempts. This standardized configuration ensures a fair and unbiased evaluation across the selected tools.

As Table \ref{tab:experiment} shows, we selected four Java projects for our experiment: Commons-Cli\cite{cli}, Commons-Csv\cite{csv}, Ecommerce-microservice\cite{ecommerce}, and Binance-connector\cite{binance}. The former two projects are widely used in related work~\cite{a3test,athenatest} and likely to be included in the training data of GPT-3.5-turbo. While the latter two are unseen popular projects (more than 100 stars) that are not part of the training data (as they are created after the creation of training data).
Considering the need for repeated operations of these IDEA plugins in the experiment, we employed random sampling with a 95\% confidence level and a 5\% margin of error. As a result, we randomly selected 264 methods from a total of 835 focal methods across these four projects. 
In our experimental procedure, we attempted to generate a test for each of these selected methods.

\begin{table}[h]
    \footnotesize
    \centering
    \caption{Result of experiment}
    \vspace*{-4mm}
    \label{tab:experiment}
    \begin{tabular}{llcccc}
    \toprule
    \multicolumn{3}{c}{Porject   Information} & \multicolumn{3}{c}{Line Coverage} \\
    \cmidrule{1-3} \cmidrule{4-6} 
    Name & Version & Unseen & ChatUniTest & TestSpark & EvoSuite \\
    \midrule
    Cli \cite{cli} & 1.5.0 & no   & 70.9\%      & 78.4\%       & \textbf{91.8\%} \\
    Csv \cite{csv} & 1.10.0 & no  & \textbf{73.3\%}     & -       & 28.3\% \\
    Ecommerce \cite{ecommerce} & 695a6d4 & yes    & 26.7\%       & \textbf{36.7\%}  &   - \\
    Binance \cite{binance} & 2.0.0 & yes   & \textbf{49.2\%}      & 29.7\%      & 20.8\% \\
    \midrule
    Overall Coverage  & -  & - & \textbf{59.6\%}      & 42.1\%      & 38.2\%  \\
    \bottomrule
    \end{tabular}
    \vspace*{-4mm}
\end{table}

In Table \ref{tab:experiment}, a comprehensive comparison of line coverage is provided for ChatUniTest, EvoSuite, and TestSpark across four projects. Notably, TestSpark fails to generate tests for the Csv project, attributed to the constructed prompt surpassing the token limit of the model. Additionally, in the Ecommerce project, EvoSuite fails to generate tests for focal methods, primarily due to JDK version incompatibility issues between EvoSuite and the Ecommerce project.

\textbf{In summary, ChatUniTest showcased relatively reliable line coverage across diverse projects, achieving an impressive line coverage of 59.6\%. This performance surpassed that of both EvoSuite and TestSpark, underscoring ChatUniTest's resilience and efficacy as a unit test generation tool.}

\paragraph{Usefulness evaluation}

We conducted a user study to investigate the usefulness of ChatUniTest. We distributed questionnaires (see~\cite{survey} for the details) to 19 individuals who either starred, forked our project, raised issues on GitHub, or reached out to us via email. Out of these, we received 9 responses.
The respondents comprised five students and four senior software developers. All participants had a foundational understanding of software testing; five of them are well-versed in the domain. Additionally, every participant had prior experience with large language models in programming.

The survey results indicate that 89\% of the respondents use ChatUniTest to assist in writing test cases. All of them believe that using ChatUniTest is beneficial for writing test cases, with 33\% of them stating that it's highly beneficial. For junior developers, especially students, ChatUniTest can effectively assist them in writing unit tests.
Moreover, 33\% of respondents are further developing upon ChatUniTest, such as adding support for additional languages, integrating new LLMs, and incorporating it into their development workflows.
Notably, some respondents provided valuable suggestions. These include integrating performance testing and supporting more testing frameworks, especially those internal testing frameworks. They also suggested measuring and verifying the reliability of the generated test code in real production environments. These suggestions not only provide guidance for ChatUniTest but also contribute to the future development of this field.
Overall, these results highlight the users' high appreciation for ChatUniTest and their expectations for its future development.

\section{conclusion and future work}
We present ChatUniTest, an automated unit test generation framework that leverages the capabilities of the  LLMs and provides a suite of user-friendly APIs to assist developers in their software testing tasks.
In the future, we plan to improve our framework to produce the optimal context and extend our preparation and validation process to support more programming languages. Moreover, we intend to provide a broader range of benchmark implementations (such as SysPrompt~\cite{SysPrompt} and ChatTester~\cite{ChatTester}), built upon the ChatUniTest.

\begin{acks}
This work was supported in part by the National Natural Science Foundation of China under Grants U20A20173 and 62125206, in part by  the National Key R\&D Program of China (2022YFF0902702), in part by  the Zhejiang Pioneer (Jianbing) Project (2023C01045), in part by the Key R\&D Program of Ningbo (2023Z235), and in part by the ZJU-Hundsun Fintech Research \& Development Center.
\end{acks}

\bibliographystyle{ACM-Reference-Format}
\bibliography{references}

\end{document}